\documentclass[ twocolumn]{IEEEtran}
\usepackage{url,amsmath,graphicx}
 \usepackage{stfloats}
 \usepackage{eqparbox}
  \usepackage{array}
  \usepackage{fixltx2e}
  \usepackage{stfloats}
   \usepackage{cite}
   \usepackage{url}
\usepackage{fancyhdr}
\usepackage{cite}
\usepackage{graphicx}
\usepackage{psfrag}
\usepackage{url}
\usepackage{stfloats}
\usepackage{amsmath}
\usepackage{array}
\usepackage{fancyhdr}

\usepackage{amssymb}
\usepackage{color}

\begin{document}
	\title{A Comprehensive Study of Reconfigurable Intelligent Surfaces in Generalized Fading }
\author{ Im\`ene~Trigui, Member, \textit{IEEE}, Wessam Ajib, Senior Member, \textit{IEEE}, and Wei-Ping Zhu, Senior Member, \textit{IEEE}. }

\maketitle
\begin{abstract}
Leveraging on the reconfigurable intelligent surface (RIS) paradigm for enabling the  next Internet of Things (IoT) and 6G era, this paper develops a comprehensive theoretical framework  characterizing the  performance of RIS-assisted communications in a plethora of propagation environments.
We derive unified mathematical models for the outage probability and ergodic capacity of single and multiple-element RIS over Fox's H fading channel, which includes as special cases nearly all linear and non linear multi-path and shadowing fading models adopted in the open literature.  For gleaning further insights,  we capitalize on
the algebraic asymptotic expansions of the H-transform to
further analyze the outage probability and capacity at high
signal-to-noise ratio (SNR) in a unified fashion.  Asymptotic analysis shows two scaling rates of the outage probability at large average SNR.
Moreover, by harnessing its tractability, the developed statistical machinery is employed to characterize the performance of multiple randomly distributed RIS-assisted communications over Fox's H fading channels.   We show that the SNR grows as $M^{\frac{\alpha}{2}} N^{2}$  when using $M$ $N$-element RISs in generalized fading channels with path-loss exponent $\alpha$.

\end{abstract}

\section{Introduction}
Contemporary wireless networks modeling and analysis is a vibrant
topic that keeps taking new dimensions in complexity as researchers relentlessly keep exploring the potential of novel breakthrough technologies to support  upcoming Internet of Things (IoT) and 6G era \cite{6G}.  Among these emerging technologies,  reconfigurable intelligent surfaces (RISs) \cite{basar1},\!\! \cite{renzo}, have been introduced  with an  overarching vision of artificially controlling the wireless environment as to increase the quality of service
and spectrum efficiency. RIS is based on massive integration of low-cost tunable passive elements  that will get weaved into conventional buildings
and objects able to transmit data by reflecting and modulating
an incident RF wave \cite{renzo}, which leads to a more controllable wireless environment. Leveraging in this key property, RIS-enabled networks challenge the device–side approaches such as massive multiple-input-multiple-output (MIMO) systems, encoding, modulation, and relaying, currently deployed in wireless networks to fully adapt to the time-variant, unpredictable channel state.  However, being fundamentally different, the RIS concept asks for new methods for modeling, analyzing, and optimizing RIS-enhanced wireless networks, yet this is still in its embryonic stage in the open literature. The so far valuable  attempts to study RIS systems include \cite{en}-\cite{Q} where several precoding optimality studies for RIS-assisted communications in terms of rate and energy efficiency  have been achieved  relying on real-time RIS phases shift control.   Recently, the interesting problems of joint active and passive beamforming  and secrecy enhancement are investigated in \cite{L1} and \cite{sec}, respectively. Even more recently,  researchers focused on outage probability \cite{out} and asymptotic data rate \cite{cap} analysis for RIS-based systems. Moreover,  the authors of  \cite{basar1}, \cite{basar2} adopted an upper bound for the average symbol error probability (SEP) harnessing on the central limit theorem (CLT) when the number of reflecting elements grows large. Unfortunately, the available results, typically providing bounds and approximations,  only consider Rayleigh fading distribution, thereby hindering the applicability of RIS-based model in setup scenarios that capture practical multi-path and shadowing conditions. Indeed, shadowing along with  high attenuation  are the main impairments at mmWave frequencies. While RIS communications are envisioned as powerful enabler for higher frequency communications,   a careful characterization of
RIS-assisted systems over composite fading conditions is crucial. Trigged by the above background, our work is pioneer in incorporating  a comprehensive multiple-parameter fading model for general case multi-path and/or shadowing
into tractable performance analysis for RIS-assisted systems without the use of CLT approximation.  On the other hand, there is a lack of literature
on the impact of the locations of multiple RISs. In \cite{tou}, the authors investigate a single-cell multiuser system aided by multiple co-located intelligent surfaces (equivalent to a single large RIS). However, the different works in \cite{basar1}-\cite{tou} consider either a single RIS or multiple RISs at given locations, and have not addressed the multi-RISs deployment issue, which
is practically important for RIS-assisted wireless systems. In this study, using BPP  to simulate the RIS's locations, we develop the first comprehensive mathematical model accounting for spatial randomness of multiple RISs in generalized fading.

Our contributions can be
summarized as follows:
\begin{itemize}
\item  We propose a novel RIS framework, where Fox's H transform theory is invoked
for modelling, in a unified fashion,  any RIS-based fading environments in terms of closed-form outage probability and  ergodic
capacity.
\item  We draw multiple useful link-level
insights from the proposed analysis. For instance, we show that the diversity gain scales with the number of elements per RIS
multiplied by the worst distribution of the fading between the
base station-RIS and RIS-user links.
\item  We study and analyze a wireless network with large-scale deployment of RISs. Under the nearest RIS association strategy, we derive the outage probability by averaging over random RIS/user distance and generalized fading.
\item We evaluate the derived outage probability and ergodic capacity expressions by simulations, and investigate
the effect of key system  and fading parameters on RIS performance and large-scale deployment.
\end{itemize}
The rest of this paper is organized as follows. Section II describes the system model and the types
of fading discussed in this paper. Then, Sections III is devoted to the  unified performance analysis framework   where the ergodic data rate and coverage probability of RIS-assisted communications  are explicitly
derived. Next, performance of large-scale RIS planning is developed in Section IV. Simulation and numerical results
are discussed in Section V and, finally, Section VI concludes the paper.

\section{SYSTEM MODEL}
	We consider  a two-dimensional RIS, composed of  $N$ tunable reflective elements, which is transmitting data to a single antenna user by reflecting an incident RF wave from a single antenna base station (BS).   The RIS can dynamically adjust the phase shift induced by
each reflecting element. Besides, we assume that the direct links from the BS to the user is
blocked by obstacles, such as buildings. Such  assumption is highly applicable in indoor
communication scenarios. Moreover, there are several works that study outdoor RIS-assisted communication
systems under a blocked direct link \cite{basar1}-\cite{basar2}. This assumption is likely to be true for
5G and beyond mmWave and sub-mmWave communication systems, which are known to suffer
from high path and penetration losses resulting in signal blockages \cite{h1}. Assuming  transmission over  flat-fading
channels, the resulting overall channel gain  is given by
\begin{equation}
h=\sqrt{\rho_L}{\bf g}{\bf \Phi}{\bf h}^{H},
\end{equation}
where $\rho_L$ is the average SNR of the RIS-assisted link, ${\bf g}=[g_1, \dots, g_N]\in {\mathbb {C}}^{1\times N}$ is the fading channel coefficients  between the BS and the RIS, ${\bf h}=[h_1, \ldots, h_N]\in \mathbb{C}^{1\times N}$ denotes the channel vector between the RIS and  the user. In addition, $
{\bf \Phi}= {\rm diag}\left(e^{j\phi_1}, e^{j\phi_2},\ldots,e^{j\phi_N} \right) \in \mathbb{C}^{N\times N}$
accounts for the effective phase shifts
applied by all RIS reflecting elements, where $\phi_n \in [0, 2\pi)$,
$n = 1, 2,\ldots,N$ are the phase-shift variables that can be optimized by the RIS.

In this paper, a general type of distribution is assumed for $g_i$ and $h_i$, $=1,\ldots, N$, as specified below.\\
\textit{Assumption 1:}   We assume that  $\left|h_i\right|$ and
$\left|g_i\right|$ are independent and non identically distributed (i.ni.d)  Fox's H-distributed RVs with respective pdf
\begin{equation}
f_{\mid y \mid_i}(x)= \kappa^{y}_{ i} {\rm H}_{p^{y}_{i},q^{y}_{i}}^{m^{y}_{i},n^{y}_{i}}\left( c^{y}_{i} x \left|
\begin{array}{ccc} (a_{ij}, A_{ij})^{y}_{j=1: p^{y}_{i}} \\ (b_{ij},B_{ij})^{y}_{j=1: q^{y}_{i}} \end{array}\right. \right),
\label{h}
\end{equation}
where $y\in\left\{h ,g\right\}$, and $H[\cdot]$ stands for the Fox'H function \cite[Eq. (1.2)]{mathai}.
The Fox's H-function pdf  considers homogeneous
radio propagation conditions and captures composite effects of multipath
fading and shadowing, subsuming large variety of extremely important or generalized fading distributions
 used in wireless communications such as Rayleigh, Nakagami-m, Weibull  $\alpha$-$\mu$, (generalized) ${\cal K}$-fading, the Fisher-Snedecor  $\cal F$, and EGK, as shown in \!\cite{FoxH} and references therein. Furthermore, the Fox's H function distribution provides enough flexibility  to account for disparate signal
	propagation mechanisms and  well-fitted to measurement data collected in
diverse propagation environments having different parameters.

\section{RIS Performance in Generalized Fading}
Using H-transforms, we now establish a unifying framework to analyze fundamental performances  for RIS-assisted wireless communication where the fading envelope is described by Fox's H distribution. The performance metrics are the outage probability and channel capacity, as well as their tradeoffs such as diversity
gain.

\subsection{Outage Probability}
\textit{Lemma 1:} For a given SNR threshold $\rho$, the outage probability in  RIS-supported network
is \begin{eqnarray}
\!\!\!\!\!\!{\bar \Pi}(\rho, N)&=& \underset{\phi_1,\ldots,\phi_n}{\max}{\rm P}\left(\log_2\left(1+ \rho_L\mid{\bf g}{\bf \Phi}{\bf h}^{H}\mid^{2} \right)<\rho\right) \nonumber \\
&=& {\rm P} \left(\log_2\left(1+ \rho_L\left(\sum_{i=1}^{N}\left| h_i\right|\left| g_i\right| \right)^{2}\right)< \rho\right).
\label{out1}
\end{eqnarray}

\textit{Proof:}
For any given ${\bf \Phi}$, the coverage expression in (\ref{out1}) is
achieved from the capacity of an additive white Gaussian noise
channel, where ${\bf g}{\bf \Phi}{\bf h}^{H}=\sum_{i=1}^{N} h_i g_i e^{j \phi_n} $.
The maximum coverage is achieved when the
phase-shifts are selected as $\phi_n=-\arg(h_n+g_n), n=1,\ldots, N$.\\

\textit{Proposition 1:}
 The outage probability achieved by RIS-assisted communication is
\begin{eqnarray}
\!\! { \Pi}(\rho, N)&=&
\tau{\rm H}_{0,1:{\widetilde p}_1, {\widetilde q}_1,\ldots, {\widetilde p}_N,  {\widetilde q}_N}^{0,0:{\widetilde m}_1, {\widetilde n}_1,\ldots, {\widetilde m}_N,  {\widetilde n}_N}\nonumber \\ && \!\!\!\!\!\!\!\!\!\!\!\!\!\!\!\!\!\!\!\!\!\!\!\!\!\!\!\!\!\!\!\!\!\!\!\!\!\!\!\!\!\left[\!\!\!\begin{array}{ccc} {\widetilde{c}}_{1} \sqrt{\rho_t}\\ \vdots \\ {\widetilde{c}}_{N} \sqrt{\rho_t}\end{array} \!\!\left |\!\!\begin{array}{ccc}  -: (1,1), (\delta_{1},\Delta_{1})_{\widetilde{p}_{1}}; \ldots; (1,1), (\delta_{N},\Delta_{N})_{\widetilde{p}_{N}} \\(0;1,\ldots, 1): (\xi_{1},\Xi_{1})_{\widetilde{q}_{1}}; \ldots;  (\xi_{N},\Xi_{N})_{\widetilde{q}_{N}} \end{array}\right. \!\!\!\!\right],
\label{pout}
\end{eqnarray}
where $\rho_t= \frac{2^{\rho}-1}{{\rho_L}}$, $\tau=\prod_{i=1}^{N}\frac{\kappa^{h}_{i}\kappa^{g}_{i}}{{c^{h}_{i}c^{g}_{i}}}$ , ${\widetilde{c}}_{i}=c^{h}_{i}c^{g}_{i}$ and $H[\cdot, \ldots, \cdot]$ is the multivariable Fox'H-function whose definition  in terms of multiple Mellin-Barnes type contour integral is given in \cite[Definition A.1]{mathai} where
\begin{equation}
(\delta_{i},\Delta_{i})_{\widetilde{p}_{i}}=\left((a_{ij}+A_{ij},A_{ij})^{h}_{j=1:p^{h}_{i}}, (a_{ij}+A_{ij},A_{ij})^{g}_{j=1:p^{g}_{i}}\right)
 \end{equation}
\begin{equation}
(\xi_{i},\Xi_{i})_{\widetilde{q}_{i}}=\left((b_{ij}+B_{ij},B_{ij})^{h}_{j=1:q^{h}_{i}}, (b_{ij}+B_{ij},B_{ij})^{g}_{j=1:q^{g}_{i}}\right)
\end{equation}
Moreover, ${\widetilde m}_i=m^{h}_{i}+ m^{g}_{i}$ , $ {\widetilde n}_i=n^{h}_{i}+ n^{g}_{i}+1$,  ${\widetilde q}_i=q^{h}_{i}+ q^{g}_{i}$, and ${\widetilde p}_i=p^{h}_{i}+ p^{g}_{i}+1$.\\
\textit{Proof:}
The probability  in (\ref{pout}) is obtained by defining the random
variables ${\cal S}=\sum_{i=1}^{N}\left|h_i\right| \left|g_i\right|$ and recognizing that
\begin{equation}
{ \Pi}(\rho, N)=\frac{1}{2 \pi j}\int_{\cal L} s^{-1}\Psi_{\cal S}(s)e^{s \sqrt{ \rho_t}}ds,
\label{eq1}
\end{equation}
where $\Psi_{\cal S}(s)=\prod_{i=1}^{N}{\cal L}(f_{\left|h_i\right|\left|g_i\right|})(s)$ where ${\cal L}(\cdot)$ stands for the Lapalce transform. After an appropriate parameter
setting in the product of two Fox' H functions using \cite[Theorem (4.1)]{FOX} then their Laplace transform by applying \cite[Eq. (2.20)]{mathai},  $\Psi_{\cal S}$ follows as shown in (\ref{mgf}) at the top of the next page.
\begin{figure*}[t!]
\begin{eqnarray}
{\Psi}_{\cal S}(s)&=&\frac{\tau}{(2\pi w)^{N}}\int_{{\cal L}_1} \ldots \int_{{\cal L}_N}\prod_{i=1}^{N}\left(\frac{\Gamma(-u_i)\Theta_i(u_i)}{{\widetilde{c}}_{i}^{u_i}}\right) s^{\sum_{i=1}^{N} u_i} du_1 du_2\ldots d_{u_N}\nonumber \\ &&
\text{where}\quad
\Theta_j(u_j)=\frac{\prod_{j=1}^{{\widetilde m}_j}\Gamma\left(\xi_{j}+\Xi_{j}u_j)\right)\prod_{j=1}^{{\widetilde n}_j}\Gamma\left(1-\delta_{j}-\Delta_{j}u_j)\right)}{\prod_{j={\widetilde n}_j+1}^{{\widetilde p}_j}\Gamma\left(\delta_j+\Delta_{j}u_i\right)\prod_{j={\widetilde m}_j+1}^{{\widetilde q}_j}\Gamma(1-\xi_{j}-\Xi_{j}u_j)}.\label{mgf}
\end{eqnarray}
\hrulefill
\end{figure*}
By plugging (\ref{mgf}) into (\ref{eq1}), the outage probability can be written as in (\ref{LMGF}), where recalling  that $\frac{1}{2 \pi j}\int_{\cal L} s^{-a} e^{s  z}ds=\frac{z^{a-1}}{\Gamma(a)}$ and  harnessing on the multiple Mellin–Barnes type contour integral of the multivariate Fox'H function \cite[Definition A.1]{mathai}, yield the desired result after some manipulations.
\begin{figure*}
	\begin{eqnarray}
\frac{1}{2 \pi j}\int_{\cal L} s^{-1}\Psi_{\cal S}(s)e^{s  z}ds&=&\frac{\tau}{(2\pi w)^{N}}\!\int_{{\cal L}_1}\ldots \int_{{\cal L}_N}\prod_{i=1}^{N}\left(\frac{\Gamma(-u_i)\Theta_i(u_i)}{{\widetilde{c}}_{i}^{u_i}}\right) \nonumber \\ && \times \frac{1}{2 \pi j}\int_{\gamma+w\infty}^{\gamma-w\infty}e^{s z} s^{\sum_{i=1}^{N} u_i-1} ds du_1 du_2\ldots d_{u_N}\nonumber \\ &=&\frac{\tau}{(2\pi w)^{N}}\!\int_{{\cal L}_1}\ldots \int_{{\cal L}_N}\prod_{i=1}^{N}\left(\frac{\Gamma(-u_i)\Theta_i(u_i)}{{\widetilde{c}}_{i}^{u_i}}\right)\frac{z^{-\sum_{i=1}^{N} u_i}}{\Gamma(1-\sum_{i=1}^{N}u_i)}du_1 du_2\ldots d_{u_N}.\label{LMGF}
	\end{eqnarray}
\hrulefill
\end{figure*}

\textit{Remark 1:} When $N=1$, the outage probability reduces to
\begin{eqnarray}
\!\!{ \Pi}(\rho, 1)&=&\tau {\mathcal H}_{{\widetilde p}_1,{\widetilde q}_1+1}^{{\widetilde m}_1,{\widetilde n}_1}\left[ {\widetilde{c}}_{1} \sqrt{\rho_t} \left|
\begin{array}{ccc} (1,1), (\delta_{1},\Delta_{1})_{\widetilde{p}_{1}} \\ (\xi_{1},\Xi_{1})_{\widetilde{q}_{1}}, (0,1) \end{array}\right. \right]\nonumber \\\!\!\!\!\!\!\!\!\!\!\!\!\!\!\!\!\!\!\!\!
&\overset{(a)}{=}&1-\kappa^{h}\kappa^{g} \sqrt{\rho_t}\nonumber \\&& \!\!\!\!\!\!\! \!\!\!\!\!\!\!\!\!\!\!\!\!\!\!\!\!\!\!\!\!\!\!\!\!\!\!\!\!\!\times {\mathcal H}_{p^{h}+p^{g}+1,q^{h}+q^{g}+1}^{m^{h}+m^{g}+1,n^{h}+n^{g}}\!\!\left[ c^{h} c^{g} \sqrt{\rho_t} \left|
\begin{array}{ccc} \!\!\!(\delta\!-\!\Delta,\Delta)_{p^{h}+p^{g}}, (0,1) \\\!\!\!(-1,1), (\xi\!-\!\Xi,\Xi)_{q^{h}+q^{g}} \end{array}\right. \!\!\!\!\right],
\label{L1}
\end{eqnarray}
where $(a)$ follows from applying \cite[Eq. (3.8)]{FOX}.\\
So far, driven by the common observation that the general case with respect to $N$ and fading distribution is rather untractable \cite{basar1}-\cite{ray},    previous works settled for only the special case when $N=1$ and Rayleigh fading. In this case,  the outage probability is obtained in \cite[Eq. (15)]{ray}. In this paper,  the special case $N=1$ specialises from the general formulas in (\ref{L1}). Moreover, in the special case of Rayleigh fading (i.e., when ${\widetilde m}=2$, ${\widetilde n}=0$, ${\widetilde p}=1$, ${\widetilde q}=2$, $\tau=1$, $ (b,B)^{h}=(b,B)^{g}=(\frac{1}{2}, \frac{1}{2})$) and  after resorting to \cite[Eq. (2.45)]{FOX} and \cite[Eq. (2.47)]{FOX}, (\ref{L1}) reduces to
\begin{equation}
{\Pi}(\rho, 1)=1-\sqrt{\rho_t} {\mathcal H}_{0,2}^{2,0}\left[  \sqrt{\rho_t} \left|
\begin{array}{ccc} - \\(-\frac{1}{2},1), (\frac{1}{2},1) \end{array}\right. \right],
\label{L2}
\end{equation}
which coincides with \cite[Eq. (15)]{ray} where the Fox's H function in (\ref{L2})
represents the Bessel function of the second kind and first order \cite[Eq. (1.128)]{mathai}.

Moreover, based on arithmetic-geometric means
inequality \cite[Sec. 11.116]{grad}, the outage probability is bounded by
\begin{eqnarray}
\!\!\!{ \Pi}(\rho, N)\!\!\!&\leq&\!\! \!{\bar \Pi}(\rho, N)={\rm P} \left(\prod_{i=1}^{N}\left| h_i\right|\left| g_i\right|< \left(\frac{\sqrt{\rho_t}}{N}\right)^{N}\right)\nonumber \\
&=&\!\!\!\!\frac{1}{2 \pi j}\int_{\cal L}\!\! s^{-1}{\cal L}(f_{\prod_{i=1}^{N}\left|h_i\right|\left|g_i\right|})(s)e^{s  \left(\frac{\sqrt{\rho_t}}{N}\right)^{N}}ds.
\end{eqnarray}
By an appropriate parameter
setting in the product of $N$ Fox' H function using \cite[Theorem (4.1)]{FOX} and applying \cite[Eq. (2.20)]{mathai} and \cite[Eq. (2.21)]{mathai} for the Laplace transform and its inverse, we obtain the outage probability upper bound as given by
\begin{eqnarray}
{\bar \Pi}(\rho, N)&=&\tau~{\rm H}_{\sum_{i=1}^{N}{\widehat p}_i+1, \sum_{i=1}^{N}{\widetilde q}_i+1}^{\sum_{i=1}^{N}{\widetilde m}_i,\sum_{i=1}^{N}{\widehat{n}}_i+1} \Bigg[\left(\frac{ \sqrt{\rho_t}}{N}\right)^{N}  \prod_{i=1}^{N}{\widetilde{c}}_{i}\nonumber \\ &&  \left|\begin{array}{ccc}   (1,1),(\delta_{j},\Delta_{j})_{{\widetilde p}_j, j=1:N} \\ (\xi_{j},\Xi{j})_{\widetilde{q}_{j}, j=1:N}, (0,1) \end{array}\right. \Bigg],
\label{poutL}
\end{eqnarray}
where $ {\widehat n}_i=n^{h}_{i}+ n^{g}_{i}$ and $ {\widehat p}_i=p^{h}_{i}+ p^{g}_{i}$. We notice that,  by virtue of  arithmetic-geometric means inequality, the multivariate Fox'H representation of the outage portability in (\ref{pout})  culminates in  a Fox' H function of a singe variable as shown in (\ref{poutL}).


\textit{Remark 2:} The main merit of theses outage probability representations is that they
rely on a versatile and generic  form of the fading distribution while accurately reflecting the behavior
of RIS networks in all operating regimes. The derived analytical expressions for the outage probabilities   in (\ref{pout}), and (\ref{poutL}) are highly generic and novel and can be easily mapped into most  existing
 fading models.  Table I lists some commonly-used channel fading distributions and
the corresponding expression for ${ \Pi}(\rho, N)$.   This
result is the first in the literature as it represents the exact SNR distribution in RIS-assisted communications
 in terms of the multivariate Fox’s H-function. This is in contrast with
the recently reported expressions in \cite[Eqs. (4), (7)]{basar1}, \cite[Eqs. (3),(9)]{basar2} and \cite[Eq. (17)]{ray}, who resorted to approximations (CLT in \cite{basar1} and moment-based Gamma approximation in \cite{ray}) to circumvent the intricacy of the exact statistical modeling of RIS-enabled communications. In this paper, the derived analytical expressions for the outage probability are obtained via the evaluation of single and multi-variable Fox's H functions.  The latters,  have been recently frequently
used in the literature and for
which efficient implementation codes exist in most popular
mathematical software packages \cite{Lei}, \cite{hmulti}. Hence, such expressions
can be very rapidly and efficiently computed.

\begin{table*}
\caption{OUTAGE  PROBABILITY OF RIS-ASSISTED COMMUNICATIONS OVER
WELL-KNOWN FADING CHANNEL MODELS}
\centering
\begin{tabular}{p{2.3in} p{4in}}
  \hline\hline
  \textbf{Instantaneous Fading Distribution} & \textbf{\quad \quad Outage Probability} $ {\Pi}(\rho, N)$ \\ \hline\hline
Nakagami-$m$ Fading \cite[Table IIV]{FoxH}:
$\begin{aligned} &f_{\mid y \mid_i}(x)=\frac{\sqrt{m^{y}_i}}{\Gamma(m^{y}_i)} {\mathcal H}_{0,1}^{1,0}\left[ \sqrt{m^{y}_i} x \left|
\begin{array}{ccc} - \\ (m^{y}_i-\frac{1}{2},\frac{1}{2}) \end{array}\right. \right]\end{aligned}$  & $\begin{aligned} \quad \quad &{\Pi}(\rho, N)=\left(\prod_{i=1}^{{N}} \Gamma(m^{h}_i)\Gamma(m^{g}_i)\right)^{-1}\\ &{\rm H}_{0,1:1,2,\ldots, 1, 2}^{0,0:2, 1,\ldots, 2, 1}\left[\!\!\!\begin{array}{ccc} \sqrt{m^{h}_1 m^{g}_1 }\sqrt{\rho_t}\\ \vdots \\ \sqrt{m^{h}_N m^{g}_N } \sqrt{\rho_t}\end{array} \!\!\left |\!\!\begin{array}{ccc}  -: (1,1), -; \ldots; (1,1), - \\(0;1,\ldots, 1): \{\xi_{1},\Xi_{1}\} ; \ldots;  \{\xi_{N},\Xi_{N}\} \end{array}\right. \!\!\!\!\right] \\ & \quad\quad \quad  \{\xi_{i},\Xi_{i}\}=(m^{h}_i, \frac{1}{2}),(m^{g}_i, \frac{1}{2}) \end{aligned}$\\\\
  \hline
$\alpha$-$\mu$ Fading \cite[Table IIV]{FoxH}:
$\begin{aligned} & f_{\mid y \mid_i}(x)=\frac{\sqrt{\eta^{y}_i}}{\Gamma(\mu^{y}_i)} {\mathcal H}_{0,1}^{1,0}\left[ \sqrt{\eta} x \left|
\begin{array}{ccc} - \\ (\mu^{y}_i-\frac{1}{\alpha^{y}_i},\frac{1}{\alpha^{y}_i}) \end{array}\right. \right]\end{aligned}$ \text{where} $\eta^{y}_i=\frac{\Gamma(\mu^{y}_i+\frac{2}{\alpha^{y}_i})}{\Gamma(\mu^{y}_i)}$
  &$\begin{aligned} \quad \quad & { \Pi}(\rho, N)=
\left(\prod_{i=1}^{{N}} \Gamma(\mu^{h}_i)\Gamma(\mu^{g}_i)\right)^{-1}\\ &{\rm H}_{0,1:1,2,\ldots, 1, 2}^{0,0:2, 1,\ldots, 2, 1}\left[\!\!\!\begin{array}{ccc} \sqrt{\eta^{h}_1 \eta^{g}_1 } \sqrt{\rho_t}\\ \vdots \\ \sqrt{\eta^{h}_N \eta^{g}_N } \sqrt{\rho_t}\end{array} \!\!\left |\!\!\begin{array}{ccc}  -: (1,1), -; \ldots; (1,1), - \\(0;1,\ldots, 1): \{\xi_{1},\Xi_{1}\}; \ldots;  \{\xi_{N},\Xi_{N}\} \end{array}\right. \!\!\!\!\right]\\ & \quad\quad \quad  \{\xi_{i},\Xi_{i}\}=(\mu^{h}_i, \frac{1}{\alpha^{h}_i}),(\mu^{g}_i, \frac{1}{\alpha^{g}_i}) \end{aligned}$\\ \\ \hline

Fisher-Snedecor $\cal F$ \cite[Eq.(3)]{fisher}:
  $\begin{aligned} & f_{\mid y \mid_i}(x)=\frac{m^{y}_i}{m^{y}_{s_i} \Gamma(m^{y}_{s_i})\Gamma(m^{y}_i)} &\\& \quad \quad \times  {\mathcal H}_{1,1}^{1,1}\left[ \frac{m^{y}_i x}{m^{y}_{s_i} }\left|
\begin{array}{ccc} (-m^{y}_{s_i}+\frac{1}{2},\frac{1}{2}) \\ (m^{y}_i-\frac{1}{2},\frac{1}{2}) \end{array}\right. \right]\end{aligned}$
& $\begin{aligned} \quad \quad &{\Pi}(\rho, N)= \left(\prod_{i=1}^{{N}} \Gamma(m^{g}_{s_i})\Gamma(m^{g}_i)\Gamma(m^{h}_{s_i})\Gamma(m^{h}_i)\right)^{-1}
\\ &{\rm H}_{0,1:3,2,\ldots, 3, 2}^{0,0:2, 3,\ldots, 2, 3}\left[\!\!\!\begin{array}{ccc} \sqrt{\frac{m^{h}_1 m^{g}_1 }{m^{h}_{s_1}m^{g}_{s_1}} } \sqrt{\rho_t} \\ \vdots \\ \sqrt{\frac{m^{h}_N m^{g}_N }{m^{h}_{s_N}m^{g}_{s_N}} } \sqrt{\rho_t}\end{array} \!\!\left |\!\!\begin{array}{ccc}  -: (1,1), \{\delta_{1},\Delta_{1}\}; \ldots; (1,1), \{\delta_{N},\Delta_{N}\}\\(0;1,\ldots, 1): \{\xi_{1},\Xi_{1}\}; \ldots;  \{\xi_{N},\Xi_{N}\} \end{array}\right. \!\!\!\!\right]\\ & \quad\quad  \{\delta_{i},\Delta_{i}\}=(1-m^{h}_{si}, \frac{1}{2}),(1-m^{g}_{si}, \frac{1}{2}) \\& \quad \quad \{\xi_{i},\Xi_{i}\}=(m^{h}_{i}, \frac{1}{2}),(m^{g}_i, \frac{1}{2})\end{aligned}$ \\\\
\hline
Generalized ${\cal K}$  \cite[Table IIV]{FoxH}:
  $\begin{aligned} & f_{\mid y \mid_i}(x)=\frac{\sqrt{m^{y}_i k^{y}_i}}{ \Gamma(m^{y}_i)\Gamma(k^{y}_i)}& \\& \!\!\!\!\! \quad \times {\mathcal H}_{0,2}^{2,0}\left[x \sqrt{m^{y}_i k^{y}_i} \left|
\begin{array}{ccc} - \\ (m^{y}_i-\frac{1}{2},\frac{1}{2}), (k^{y}_i-\frac{1}{2},\frac{1}{2}) \end{array}\right. \right]\end{aligned}$
& $\begin{aligned} \quad \quad &{\Pi}(\rho, N)=
\left(\prod_{i=1}^{{N}} \Gamma(m^{h}_i)\Gamma(\kappa^{h}_i)\Gamma(m^{g}_i)\Gamma(\kappa^{g}_i)\right)^{-1}\\ &{\rm H}_{0,1:1,4,\ldots, 1, 4}^{0,0:4, 1,\ldots, 4, 1}\left[\!\!\!\!\!\begin{array}{ccc} \sqrt{m^{h}_1 \kappa^{h}_1 m^{g}_1 \kappa^{g}_1 } \sqrt{\rho_t}\\ \vdots \\ \sqrt{m^{h}_N \kappa^{h}_N m^{g}_N \kappa^{g}_N} \sqrt{\rho_t}\end{array} \!\!\!\!\left |\!\!\begin{array}{ccc}  -: (1,1), -; \ldots; (1,1), -\\(0;1,\ldots, 1): (\xi_{1},\Xi_{1})_{p_{11}}; \ldots;  (\xi_{N},\Xi_{N})_{p_{N}} \end{array}\right. \!\!\!\!\right]\\ & \quad\quad \quad  \{\xi_{i},\Xi_{i}\}=(m^{h}_i, \frac{1}{2}),(m^{g}_i, \frac{1}{2}),(k^{h}_i, \frac{1}{2}),(k^{g}_i, \frac{1}{2})\end{aligned}$ \\\\
\hline
\end{tabular}
\end{table*}

\subsection{Diversity Analysis}
In an effort to understand the impact of some key system
parameters on the outage probability, we analyze the asymptotic regime at  high SNR from which we derive the diversity and coding gains. Asymptotic analysis is particulary useful to this framework since the evaluation of the multivariate Fox's H function may encounter underflow problems  when $N$ is large.

\textit{Proposition 2:} The asymptotic expansion of the outage probability ${\bar \Pi}(\rho, N)$ in (\ref{pout}) for high SNRs can be obtained by computing the residue \cite{kilbas}. Let us consider the residue at the points $\zeta_l = (\zeta_1, \ldots, \zeta_N)$, where $\zeta_l=\underset{j=1,\ldots, \widetilde{m_l}}{\min}\left\{\frac{\xi_{lj}}{\Xi_{lj}}\right\}$,  the asymptotic outage probability\footnote{It should be stressed here that (\ref{asymout}) holds when  poles at $\{\zeta_1, \ldots, \zeta_N \}$ are simple, i.e. $\Xi_{ti}(\xi_{lj}+k)\neq \Xi_{lj}(\xi_{ti}+k'), i\neq j, i,j=1,\ldots,\widetilde{m_l},  l=0,1,\ldots,N, k,k'=0,1,\ldots,N$ \cite{kilbas}, which is particulary verified in i.ni.d fading.} is obtained as
\begin{equation}
{ \Pi}(\rho, N)\approx\frac{\tau}{\Gamma(1+\sum_{i=1}^{N}\zeta_i)}\left(\!\prod_{i=1}^{N} \widetilde{\Theta}_i(-\zeta_i){\widetilde c}_i^{\zeta_i}\!\!\right)\!\rho_t^{\frac{\sum_{i=1}^{N}\zeta_i}{2}},
\label{asymout}
\end{equation}
 where
 $ \widetilde{\Theta}_j(\zeta_j)=$ $\frac{\prod_{i=1, i\neq j}^{{\widetilde m}_j}\Gamma\left(\xi_{i}+\Xi_{i}\zeta_j\right)\prod_{i=1}^{{\widetilde n}_j}\Gamma\left(1-\delta_{i}-\Delta_{i}\zeta_j)\right)}{\prod_{i={\widetilde n}_j+1}^{{\widetilde p}_j}\Gamma\left(\delta_i+\Delta_{i}\zeta_j\right)\prod_{i={\widetilde m}_j+1}^{{\widetilde q}_j}\Gamma(1-\xi_{i}-\Xi_{i}\zeta_j)}$.


From (\ref{asymout}), the diversity order of the considered RIS-assisted system is given by
\begin{equation}
{\cal G}_d=\lim_{\rho_L\rightarrow\infty}\frac{\log {\bar \Pi}(\rho, N) }{\log \rho_L}=\frac{\sum_{i=1}^{N}\zeta_i}{2}.
\end{equation}
On the other hand, exploiting the  outage probability upper bound in (\ref{poutL}),  the performance trends of RIS-assisted communications scales as
\begin{equation}
{ \Pi}(\rho, N)\overset{(a)}{\leq}\tau h^{*}\left(\left(\frac{ \sqrt{\rho_t}}{N}\right)^{N} \prod_{i=1}^{N}{\widetilde{c}}_{i}\right)^{\underset{l=1,\ldots, \sum_{i=1}^{N}\widetilde{m_l}}{\min}\zeta_l},
\label{outlb}
\end{equation}
where $(a)$ follows from applying the asymptotic expansion of the Fox's H function in \cite[Eq. (1.94)]{mathai}, where $h^{*}$ follows from an appropriate parameters setting in \cite[Eq. (1.5.8)]{kilbas}.\\
It can be inferred from (\ref{outlb}) that the SNR in RIS-assisted communication may grow quadratically with $N$, which is consistent with recent works in the RIS literature \cite{basar2}-\cite{bor}. Yet, the latters  did not explain how this SNR gain could affect the outage performance in generalized channel models.  The $N^{2}$ array gain (\ref{outlb}) can be interpreted by an augmented received power by the RIS due to $N$ reflecting elements multiplied by the power/array gain achieved by  passive beamforming  when having a large array.
 
 Under the same rationale of (\ref{pout}) and its asymptotic expansion in (\ref{asymout}) while applying the arithmetic-quadratic inequality, i.e. $\left(\sum_{i=1}^{N}\left|h_i\right|\left|g_i\right|\right)^{2}\leq N \sum_{i=1}^{N}h_i^{2}g_i^{2}$, we obtain 
\begin{equation}
{ \Pi}(\rho, N)\overset{(a)}{\geq} {\cal \widetilde{C}} \left(\frac{ \rho_t}{N}\right)^{\sum_{i=1}^{N}\widetilde{\zeta}_i},
\label{outlb2}
\end{equation}
where $\widetilde{\zeta}_i=2 \zeta_i$, due to the distribution of $ h_i^{2}g_i^{2}$  after using the Fox's H function property ${\rm H}_{p,q}^{m,n}\big[x \big|
\begin{array}{ccc} (a_i, k A_j)_p \\ (b_i, k B_j)_q \end{array}\big. \big]=\frac{1}{k}{\rm H}_{p,q}^{m,n}\big[x^{\frac{1}{k}} \big|
\begin{array}{ccc} (a_i, A_j)_p \\ (b_i, B_j)_q \end{array}\big. \big]$, $k>0$ \cite{mathai}. Moreover $(a)$ follwos from  a direct application of   \cite[Eq. (1.5.8)]{kilbas}.\\
Interestingly, the lower bound in (\ref{outlb2})  shows that RIS-assited SNR growth is faster than the linear scaling with $N$ observed for massive MIMO receiver in \cite{L1}, \cite{imen2} and for the MIMO relay in \cite{bor}. It should be stressed here that the upper and lower bounds  in (\ref{outlb}) and (\ref{outlb2}) contains the product of the total source-RIS and RIS-user channel gains. This is defienitely the structure of the far-field scenario where the links distances appears within seperate multiplicative fucntions \cite{basar2}-\cite{ray}. The theoretical performance limits of RIS-assited communications in near-field  senarios  and general fading models is still an open problem \cite{bor}.

 Hereafter, we specify the asymptotic performance of RIS-assisted communications in small-scale fading (ex. Nakagami-$m$) and composite small-scale/shadowing fading (ex. generalized $\cal K$) channels.

\textit{Corollary 2:} The asymptotic outage probability of RIS-assisted communications in Nakagami-$m$ fading is
\begin{eqnarray}
{ \Pi}(\rho, N)&\approx& \tau {\cal C}  \rho_L^{-\sum_{i=1}^{N}\min\{m^{g}_i,m^{h}_i\}},
\label{ounak}
\end{eqnarray}
 where $\rho_L$ stands for the average SNR, $\tau=\left(\prod_{i=1}^{{N}} \Gamma(m^{h}_i)\Gamma(m^{g}_i)\right)^{-1}$ and  ${\cal C}=\frac{\prod_{i=1}^{N} \Gamma\left(m^{g}_i-\frac{\zeta_i}{2}\right)\Gamma\left(m^{h}_i-\frac{\zeta_i}{2}\right)\prod_{i=1}^{N}(m^{g}_i m^{h}_i)^{\min\{m^{g}_i,m^{h}_i\}}}{\Gamma\left(1+2\sum_{i=1}^{N}\min\{m^{g}_i,m^{h}_i\}\right)}$. Moreover, resorting to the asymptotic upper bound in (\ref{outlb}), it follows that
\begin{equation}
{\bar \Pi}(\rho, N)\approx \tau {\bar {\cal C}} \left(\frac{ \rho_t}{N^{2}}\right)^{N \min \{m^{g}_1, m^{h}_1, \ldots,m^{g}_N, m^{h}_N \}},
\label{outlbnak}
\end{equation}
where ${\bar {\cal C}}=\frac{ \left(\prod_{i=1}^{N}(m^{g}_i m^{h}_i)\right)^{{\min \{m^{g}_1, m^{h}_1, \ldots,m^{g}_N, m^{h}_N }\}}}{\prod_{i=1}^{{N}} \Gamma(m^{h}_i)\Gamma(m^{g}_i)} $.\\

Corollary 2 stipulates that RIS performance degrades when the propagation environment
exhibits poor scattering conditions (smaller $m^{h}$ and $m^{g}$).  However, at higher frequencies (mmWave and sub-mmWave), the propagation conditions get harsher since mmWave signals are extremely sensitive to objects, including foliage and human body, resulting in signal blockage. Recently, the authors of \cite{triguifso}  studied the generalised $\cal K$ to provides accurate
modeling and characterisation of the simultaneous occurrence
of multipath fading and shadowing in mmWave communications. In RIS context we obtain the following Corollary.\\
\textit{Corollary 3:} The asymptotic outage probability of RIS-assisted communications in Generalised $\cal K$ fading is
\begin{eqnarray}
{ \Pi}(\rho, N)&\approx& \tau {\cal C}  \rho_L^{-\sum_{i=1}^{N}\min\{\kappa^{g}_i, m^{g}_i, \kappa^{h}_i, m^{h}_i\}},
\end{eqnarray}
where $\tau$  and $\cal C$ follows under the same rational of (\ref{ounak}) while using the fourth line in Table I.
Corollary 2 reveals that the diversity gain of RIS-assisted communications in composite multi-path shadowing fading channels  is limited to the worst channel condition
between multipath fading and shadowing on both BS-RIS and RIS-user links.

\textit{Corollary 4 (Outage scaling in i.i.d. Fox's H fading):}
When $\left|h_i\right|$ and
$\left|g_i\right|$ are i.i.d. Fox's H-distributed RVs, then a power-logarithmic series expansion of the outage probability is given by
\begin{equation}
{\bar \Pi}(\rho, N)\overset{(a)}{\approx}  {\cal T}  \rho_t^{ \frac{N\xi}{2\Xi}}\ln\left(\rho_t^{-1}\right)^{2 N m-1},
\label{outlb1}
\end{equation}
where $(a)$ follows from applying \cite[Eq. (1.4.18)]{kilbas} with ${\cal T}=H^{*}\left(\frac{\kappa}{c}\right)^{2N}  \left(  \frac{c^{N}}{N^{2}}\right)^{N \frac{\xi}{\Xi}}$, where $H^{*}$ is given in \cite[Eq. (1.4.6)]{kilbas}.  The above result shows that high SNR outage of RIS-based communications in i.i.d. fading scales as $\rho_L^{-N \frac{\xi}{2\Xi}}\ln\left(\rho_L\right)^{2Nm-1}$, where $\rho_L$ is the average transmit SNR. Specifically, under i.i.d. Nakagami-$m$ fading we obtain
\begin{equation}
{\bar \Pi}(\rho, N)\approx {\cal T}  \rho_t^{N m}\ln(\rho_t^{-1})^{2 N-1},
\label{outlb1}
\end{equation}
implying that the outage scales as $\rho_L^{-N} \ln(\rho_L)^{2N-1}$ in Rayleigh fading. Previously the authors of \cite{ray} reveled that the outage decreases at the rate $\rho_L^{-N} \ln(\rho_L)^{N}$, which coincides with (\ref{outlb1}) when $N=1$.



\textit{Remark 3 (A hint on RIS-assisted optical communications):}
The concept of using RISs in free space optical (FSO) links is viable as  to relax the line of-sight requirement of FSO systems.  Lately, in
\cite{FSO}, the concept of using RISs in FSO links was presented
as a cost-effective solution for backhauling of cellular systems.
However, the focus of \cite{FSO} was on network planning and the
impact of RISs on the FSO channel model was not studied. Recently, the authors of \cite{fisher}, \cite{fisher2} proposed  the Fischer-Snedecor $\cal F$-distribution, for which the RIS framework is obtained in Table I, for modeling turbulence-induced fading in free-space optical systems. In \cite{fisher}, \cite{fisher2} the small-scale irradiance variations of the propagating wave are modeled by a gamma distribution, while the  large-scale irradiance fluctuations follow  an inverse gamma distribution.
For this new turbulence distribution, the small- and large-scale irradiance variances and henceforth $m$ and $m_s$  in [line 3, Table I] are expressed in terms of important parameters affecting optical propagation, including the atmospheric refractive-index structure parameter, the propagation path length, the inner and the outer scale of turbulence as shown in \cite[Eqs. (13),(14), (16), (17)]{fisher2}.

\textit{Remark 4:} Corollary 2,3, and 4 demonstrate the unification of various FSO turbulent and RF fading scenarios into a single closed-form expression  for RISs performance. More importantly, capitalizing on the versatility of the Fox's H distribution and the generality of the Fox'H transform theory, the framework of this paper  provides a  powerful baseline model to
build upon to potentially extend the results of this
paper to many other directions. Without any pretention of
being able to discuss them all due to lack of space, the
most prominent directions for future works include MIMO and
cellular networks relying on RISs. Interestingly, the proposed framework not only promotes general generic fading channels, but also other generalization aspects in terms of path-loss models (see Section IV), LOS/NLOS propagation and random blockage.

\subsection{Channel Capacity}
\textit{Proposition 3:}
 The RIS-assisted communication channel capacity defined as
\begin{eqnarray}
{\cal E}(\rho_L,N) &\triangleq&\frac{ E\left\{\ln\left(1+ \rho_L \left(\sum_{i=1}^{N}\left| h_i\right|\left| g_i\right| \right)^{2}\right)\right\}}{\ln(2)},
\label{cap}
\end{eqnarray}
is obtained as in (\ref{capf}) at the top of the next page.
\begin{figure*}
\begin{eqnarray}
{\cal E}(\rho_L,N)&=&\frac{\tau}{\ln(2)(2\pi w)^{N}}\!\int_{{\cal L}_1}\ldots \int_{{\cal L}_N}\frac{\prod_{i=1}^{N}\left(\frac{\Gamma(-u_i)\Theta_i(u_i)}{{\widetilde{c}}_{i}^{u_i}}\right)}{\Gamma(1-\sum_{i=1}^{N}u_i)} \int^{\infty}_{0} E_{i}(-s) \int_{0}^{\infty} z^{1-\sum_{i=1}^{N}u_i} e^{-\rho_L s z^{2}} dz ~ds ~du_1 du_2\ldots d_{u_N}\nonumber \\
&=&\frac{\tau}{\ln(2)(2\pi w)^{N}}\!\int_{{\cal L}_1}\ldots \int_{{\cal L}_N}\frac{\prod_{i=1}^{N}\left(\frac{\Gamma(-u_i)\Theta_i(u_i)}{\left(\sqrt{\frac{1}{\rho_L}}{\widetilde{c}}_{i}\right)^{u_i}}\right)}{\Gamma\left(1-\sum_{i=1}^{N}u_i\right)}\frac{
\Gamma\left(\frac{\sum_{i=1}^{N}u_i}{2}\right)}{\Gamma\left(1-\frac{\sum_{i=1}^{N}u_i}{2}\right)}du_1 du_2\ldots d_{u_N}.
\label{capi}
\end{eqnarray}
\hrulefill
\end{figure*}
\begin{figure*}
\begin{eqnarray}
\!\!{\cal E}(\rho_L,N)\!\!\!\!&=&\!\!\!\!\frac{\tau}{\ln(2)}\nonumber \\ && \!\!\!\!\!\!\!\!\!\!\!\!\!\!\times 
{\rm H}_{2,1:{\widetilde p}_1, {\widetilde q}_1,\ldots, {\widetilde p}_N,  {\widetilde q}_N}^{0,1:{\widetilde m}_1, {\widetilde n}_1,\ldots, {\widetilde m}_N,  {\widetilde n}_N}\!\!\left[\!\!\!\begin{array}{ccc} \frac{{\widetilde{c}}_{1}}{\sqrt{\rho_L}}\\ \vdots \\ \frac{ {\widetilde{c}}_{N}}{\sqrt{\rho_L}} \end{array} \!\!\left |\!\!\begin{array}{ccc}  (1;-\frac{1}{2},\ldots,-\frac{1}{2}), (1;-\frac{1}{2},\ldots,-\frac{1}{2}):(1,1),(\delta_{1},\Delta_{1})_{p_{1}}; \ldots; (1,1), (\delta_{N},\Delta_{N})_{p_{N}} \\(1;-1,\ldots, -1): (\xi_{1},\Xi_{1})_{p_{11}}; \ldots;  (\xi_{N},\Xi_{N})_{p_{N}}\! \end{array}\right. \!\!\!\!\right].
\label{capf}
\end{eqnarray}
\hrulefill
\end{figure*}

\textit{Proof:}
The ergodic  capacity can be expressed as
\begin{equation}
{\cal E}=\int_{0}^{\infty} E_i(-x) \frac{\partial\Psi_{{\cal S}^{2}}(\rho_L x)}{\partial x} dx,
\label{capi}
\end{equation}
where  $E_i(\cdot)$ stands for the exponential integral function \cite{grad}. Moreover, referring to the relation with the MGF $\Psi_{{\cal S}^{2}}(s)$ and its derivative, i.e $\frac{\partial\Psi_{{\cal S}^{2}}(s)}{\partial s}=-\mathbb{E}\{{\cal S}^{2}e^{-s{\cal S}^{2}}\}=-\int^{\infty}_{0} x^{2} e^{-s x^{2}} f_{\cal S}(x)dx$,  where $f_{\cal S}(x)$ is the PDF of ${\cal S}$ obtained from differentiating (\ref{LMGF}). Plugging all together, as shown in (\ref{capi}) at the top of the next page,  then  applying \cite[Eq. (3.478)]{grad} and  \cite[Eq. (6.223)]{grad} and recalling the Mellin barnes integral representation of the multivariable Fox' H function \cite[Definition A.1]{mathai}, yields the desired result after several manipulations.


\textit{Remark 5:}
When $N=1$, then with the aid of (\ref{L1}) and  recalling that ${\cal E}=\frac{1}{\ln(2)}\int_{0}^{\infty}\frac{1-{\Pi}(x, 1)}{1+x}dx$,   the ergodic capacity is expressed  as
\begin{eqnarray}
\!\!{\cal E}(\rho_L,1)&=&\frac{\kappa^{h}\kappa^{g} }{\ln(2)\sqrt{\rho_L}}\int_{0}^{\infty}\sqrt{x}{\rm H}_{1,1}^{1,1}\left[x \left|
\begin{array}{ccc} (0,1) \\(0,1) \end{array}\right. \right]\nonumber \\&&\!\!\!\!\!\!\!\!\!\!\!\!\!\!\!\!\!\!\!\!\!\!\!\!\!\!\!\!\!\!\!\!\!\!\!\!\!\!{\rm H}_{{\widetilde p}_1,{\widetilde q}_1+1}^{{\widetilde m}_1+1,{\widetilde n}_1-1}\!\!\left[ {\widetilde{c}}_{1} \sqrt{\frac{x}{\rho_L}} \left|
\begin{array}{ccc} (\delta_{1}\!-\!\Delta_1,\Delta_{1})_{\widetilde{p}_{1}}, (0,1) \\(-1,1), (\xi_{1}-\Xi_1,\Xi_{1})_{\widetilde{q}_{1}} \end{array}\right. \right]dx,
\end{eqnarray}
which after applying \cite[Eq. (2.3)]{mathai} and \cite[Eq. (1.60)]{mathai} yields
\begin{eqnarray}
{\cal E}(\rho_L,1)&=&\frac{\tau}{\ln(2)}\nonumber \\ && \!\!\!\!\!\!\!\!\!\!\!\!\!\!\!\!\!\!\!\!\!\!\!\!\!\!\!\!{\rm H}_{{\widetilde p}_1+1,{\widetilde q}_1+2}^{{\widetilde m}_1+2,{\widetilde n}_1}\left[ \frac{{\widetilde{c}}_{1}}{\sqrt{\rho_L}} \left|
\begin{array}{ccc} (0,\frac{1}{2}),(\delta_{1},\Delta_{1})_{\widetilde{p}_{1}}, (1,1) \\(0,1), (0,\frac{1}{2}), (\xi_{1},\Xi_{1})_{\widetilde{q}_{1}} \end{array}\right. \right].
\label{cap1}
\end{eqnarray}
While the above result  reduces to \cite[Eq. (24)]{ray} in Rayleigh fading, it unifies the performance of single-element RIS-enabled communications in a plethora of  fading channels stemming from the versatile Fox's H fading model.

When $N\geq2$, a lower bound on the ergodic capacity is obtained from (\ref{poutL}) after following similar steps as in (\ref{cap1}, thereby yielding
\begin{eqnarray}
{\cal E}(\rho_L,N)\geq\frac{\tau}{\ln(2)}~ {\rm H}_{\sum_{i=1}^{N}{\widehat p}_i+2, \sum_{i=1}^{N}{\widetilde q}_i+2}^{\sum_{i=1}^{N}{\widetilde m}_i+2,\sum_{i=1}^{N}{\widehat{n}}_i+1}\nonumber \\ &&\!\!\!\!\!\!\!\!\!\!\!\!\!\!\!\!\!\!\!\!\!\!\!\!\!\!\!\!\!\!\!\!\!\!\!\!\!\!\!\!\!\!\!\!\!\!\!\!\!\!\!\!\!\!\!\!\!\!\!\!\!\!\!\!\!\!\!\!\!\!\!\!\!\!\!\!\!\!\!\!\!\!\!\!\!\!\!\!\!\!\!\!\!\!\!\!\!\!\!\!\!\!\!\!\left[\!\frac{ \prod_{i=1}^{N}{\widetilde{c}}_{i} }{\left(N\sqrt{\rho_L}\right)^{N}} \! \left|\!\begin{array}{ccc}   (0,\frac{N}{2}),(\delta_{j},\Delta_{j})_{{\widetilde p}_j, j=1:N}, (1,1) \\ (0,1), (0,\frac{N}{2}),(\xi_{j},\Xi{j})_{\widetilde{m}_{j}, j=1:\widetilde{q}_{j}} \end{array}\right.\right].
\label{cL}
\end{eqnarray}
For high SNR, i.e. $\rho_L\gg1$,  and  relying on the same rationale leading to  Proposition 2, (\ref{outlb}) and (\ref{outlb1}), we show that the ergodic capacity of $N$ elements RIS-assisted communications increases with the rate
$\ln(N^{2}\rho_L)$. Indeed, under the assumption  that $\Xi_{ti}(\xi_{lj}+k)\neq \Xi_{lj}(\xi_{ti}+k'), i\neq j, i,j=1,\ldots,\widetilde{m_l}, l\neq t,  l,t=0,1,\ldots,N, k,k'=0,1,2, \ldots$, then only the  poles of $\Gamma(s)$ and $\Gamma(N s/2)$ in (\ref{poutL}) coincide with order $2$. Moreover, since $\min\{0, \zeta_l\}=0, l=1,\ldots, N$, then according to \cite[Eq. (1.8.13)]{kilbas} a scaling rate of $\ln(N^{2}\rho_L)$ is obtained in the high SNR regime.

\begin{table*}
\caption{CHANNEL CAPACITY OF RIS-ASSISTED COMMUNICATIONS OVER
WELL-KNOWN FADING CHANNEL MODELS}
\centering
\begin{tabular}{p{2in} p{4.8in}}
  \hline\hline
  \textbf{Instantaneous Fading Distribution} & \textbf{\quad \quad Channel Capacity} $ {\cal E}(\rho_L,N)$ \\ \hline\hline
Nakagami-$m$ Fading \cite[Table IIV]{FoxH}:
  & $\begin{aligned} \quad \quad &{\cal E}(\rho_L,N)=\frac{\left(\prod_{i=1}^{{N}} \Gamma(m^{h}_i)\Gamma(m^{g}_i)\right)^{-1}}{\ln(2)}\\ &{\rm H}_{2,1:1,2,\ldots, 1, 2}^{0,1:2, 1,\ldots, 2, 1}\left[\!\!\!\begin{array}{ccc} \frac{\sqrt{m^{h}_1 m^{g}_1 }}{\sqrt{\rho_L}}\\ \vdots \\ \frac{\sqrt{m^{h}_N m^{g}_N } }{\sqrt{\rho_L}}\end{array} \!\!\left |\!\!\begin{array}{ccc} (1;\{-\frac{1}{2}\}_{1:N}), (1;\{-\frac{1}{2}\}_{1:N}): (1,1), -; \ldots; (1,1), - \\(1;1,\ldots, 1): \{\xi_{1},\Xi_{1}\} ; \ldots;  \{\xi_{N},\Xi_{N}\} \end{array}\right. \!\!\!\!\right] \\ & \quad\quad \quad  \{\xi_{i},\Xi_{i}\}=(m^{h}_i, \frac{1}{2}),(m^{g}_i, \frac{1}{2}) \end{aligned}$\\\\
  \hline
$\alpha$-$\mu$ Fading \cite[Table IIV]{FoxH}:
  &$\begin{aligned} \quad \quad & {\cal E}(\rho_L,N)=\frac{
\left(\prod_{i=1}^{{N}} \Gamma(\mu^{h}_i)\Gamma(\mu^{g}_i)\right)^{-1}}{\ln(2)}\\ &{\rm H}_{2,1:1,2,\ldots, 1, 2}^{0,1:2, 1,\ldots, 2, 1}\left[\!\!\!\begin{array}{ccc} \frac{\sqrt{\eta^{h}_1 \eta^{g}_1 }}{\sqrt{ \rho_L}}\\ \vdots \\ \frac{\sqrt{\eta^{h}_1 \eta^{g}_1 }}{ \sqrt{\rho_L}}\end{array} \!\!\left |\!\!\begin{array}{ccc}  (1;\{-\frac{1}{2}\}_{1:N}), (1;\{-\frac{1}{2}\}_{1:N}): (1,1), -; \ldots; (1,1), - \\(1;1,\ldots, 1): \{\xi_{1},\Xi_{1}\}; \ldots;  \{\xi_{N},\Xi_{N}\} \end{array}\right. \!\!\!\!\right]\\ & \quad\quad \quad  \{\xi_{i},\Xi_{i}\}=(\mu^{h}_i, \frac{1}{\alpha^{h}_i}),(\mu^{g}_i, \frac{1}{\alpha^{g}_i}) \end{aligned}$\\ \\ \hline

Fisher-Snedecor $\cal F$ \cite[Eq.(3)]{fisher}:
  & $\begin{aligned} \quad \quad &{\cal E}(\rho_L,N)= \frac{\left(\prod_{i=1}^{{N}} \Gamma(m^{g}_{s_i})\Gamma(m^{g}_i)\Gamma(m^{h}_{s_i})\Gamma(m^{h}_i)\right)^{-1}}{\ln(2)}
\\ &{\rm H}_{2,1:3,2,\ldots, 3, 2}^{0,1:2, 3,\ldots, 2, 3}\left[\!\!\!\begin{array}{ccc}\frac{\sqrt{\frac{m^{h}_1 m^{g}_1 }{m^{h}_{s_1}m^{g}_{s_1}} }}{\sqrt{\rho_L}}\\ \vdots \\ \frac{\sqrt{\frac{m^{h}_N m^{g}_N }{m^{h}_{s_N}m^{g}_{s_N}} }}{\sqrt{\rho_L}}\end{array} \!\!\left |\!\!\begin{array}{ccc} (1;\{-\frac{1}{2}\}_{1:N}), (1;\{-\frac{1}{2}\}_{1:N}): (1,1), \{\delta_{1},\Delta_{1}\}; \ldots; (1,1), \{\delta_{N},\Delta_{N}\}\\(1;1,\ldots, 1): \{\xi_{1},\Xi_{1}\}; \ldots;  \{\xi_{N},\Xi_{N}\} \end{array}\right. \!\!\!\!\right]\\ & \quad\quad  \{\delta_{i},\Delta_{i}\}=(1-m^{h}_{si}, \frac{1}{2}),(1-m^{g}_{si}, \frac{1}{2}) \\& \quad \quad \{\xi_{i},\Xi_{i}\}=(m^{h}_{i}, \frac{1}{2}),(m^{g}_i, \frac{1}{2})\end{aligned}$ \\\\
\hline
Generalized ${\cal K}$  \cite[Table IIV]{FoxH}:

& $\begin{aligned} \quad \quad &{\cal E}(\rho_L,N)=\frac{
\left(\prod_{i=1}^{{N}} \Gamma(m^{h}_i)\Gamma(\kappa^{h}_i)\Gamma(m^{g}_i)\Gamma(\kappa^{g}_i)\right)^{-1}}{\ln(2)}\\ &{\rm H}_{2,1:1,4,\ldots, 1, 4}^{0,1:4, 1,\ldots, 4, 1}\left[\!\!\!\begin{array}{ccc}\frac{ \sqrt{m^{h}_1 \kappa^{h}_1 m^{g}_1 \kappa^{g}_1}}{\sqrt{\rho_L}}\\ \vdots \\ \frac{\sqrt{m^{h}_N \kappa^{h}_N m^{g}_N \kappa^{g}_N}}{ \sqrt{\rho_L}}\end{array} \!\!\left |\!\!\begin{array}{ccc}  (1;\{-\frac{1}{2}\}_{1:N}), (1;\{-\frac{1}{2}\}_{1:N}): (1,1), -; \ldots; (1,1), -\\(1;1,\ldots, 1): (\xi_{1},\Xi_{1})_{p_{11}}; \ldots;  (\xi_{N},\Xi_{N})_{p_{N}} \end{array}\right. \!\!\!\!\right]\\ & \quad\quad \quad  \{\xi_{i},\Xi_{i}\}=(m^{h}_i, \frac{1}{2}),(m^{g}_i, \frac{1}{2}),(k^{h}_i, \frac{1}{2}),(k^{g}_i, \frac{1}{2})\end{aligned}$ \\\\
\hline
\end{tabular}
\end{table*}

\section{Performance of Large-Scale RIS Deployment}
In order to evaluate the efficiency of large scale deployment of RIS, it is curial to select an appropriate model for the path-loss experienced by the received signal through  RIS-assisted communications. Recently, the authors of \cite{path} stipulated that the transmitter-RIS-path loss scales as $N^{2}(d+r)^{-\alpha}$, where $d$ and $r$ are the BS-IRS, and IRS-receiver distances, respectively, and $\alpha$ is the path-loss exponent.
 In what follows,  we consider a  homogeneous binomial point process (BPP) \cite{dist} $\Phi$, with $M$
transmitting RISs uniformly distributed  in the distance range $[0,R]$ form the user.   Assuming that each receiver can be connected to its nearest RIS and denoting $\widetilde{{\Pi}}(\rho, N)=\mathbb{E}\left\{{\Pi}(\rho, N)\right\}$ as the spatial outage averaged
over the distribution of the RISs random locations, then the outage probability of a typical receiver located at the origin is given by 
\begin{equation}
\widetilde{{\Pi}}(\rho, N)=\int_{0}^{R} {\cal L}^{-1}\left\{\frac{\Psi(s)}{s},\frac{\sqrt{\rho_t}}{(d+x)^{-\frac{\alpha}{2}}} \right\}f_r(x) dx.
\label{pim}
\end{equation}
 where the $f_r(x)$ is the nearest neighbor distance pdf in BPP given by \cite{dist}
\begin{equation}
f_r(x)=\frac{2 M}{x} \left(1-\left(\frac{x}{R}\right)^{2}\right)^{M-1}\left(\frac{x}{R}\right)^{2}, \quad 0<x<R.
\label{fr}
\end{equation}
For sake of a tractable solution to (\ref{pim}), we assume  that at least one of the transmitters and the receivers is in the surface far-field. While such an assumption is more applicable in indoor
communication scenarios with users close to the walls and every wall covered by an IRS (far-field transmitter),  Another potential case  is to use the IRS as a MIMO transmitter where a single-antenna transmitter near an IRS can be jointly configured to act as a MIMO beamforming array (far-filed receiver). Accordingly, assuming that $d\gg r$, we have $(d+r)^{-\alpha/2}\simeq d^{-\alpha/2}r^{-\alpha/2}$. Using such 
separate multiplicative representation of distances recall for the previously analyzed outage scaling rates in (\ref{outlb}) and (\ref{outlb2}).
Subsequently, using (\ref{pim}), we obtain the outage probability expression as shown at the top of the next page,
\begin{figure*}
\begin{eqnarray}
\!\! \widetilde{{\Pi}}(\rho, N)\!\!\!&=& \!\!\! 2 M \tau\!\!\sum_{k=0}^{M-1}\!\!\binom{\!M\!-\!1\!}{k}\!(-1)^{k}
{\rm H}_{0,2:{\widetilde p}_1, {\widetilde q}_1,\ldots, {\widetilde p}_N,  {\widetilde q}_N}^{0,1:{\widetilde m}_1, {\widetilde n}_1,\ldots, {\widetilde m}_N,  {\widetilde n}_N}\left[\!\!\!\begin{array}{ccc} \frac{{\widetilde{c}}_{1} R ^{\frac{\alpha}{2}}\sqrt{\rho_t}}{d^{-\frac{\alpha}{2}}} \\ \vdots \\ \frac{{\widetilde{c}}_{N} R ^{\frac{\alpha}{2}} \sqrt{\rho_t}}{ d^{-\frac{\alpha}{2}}} \end{array} \left |\begin{array}{ccc} (-1-2k; \frac{\alpha}{2}, \ldots, \frac{\alpha}{2}): {\cal A}\\(0;1,\ldots, 1), (-2-2k; \frac{\alpha}{2}, \ldots, \frac{\alpha}{2}): {\cal B}\end{array}\right. \!\!\!\!\right],
\label{poutd}
\end{eqnarray}
\hrulefill
\end{figure*}
where ${\cal A}=(1,1), (\delta_{1},\Delta_{1})_{\widetilde{p}_{1}}; \ldots; (1,1), (\delta_{N},\Delta_{N})_{\widetilde{p}_{N}}$ and $ {\cal B}= (\xi_{1},\Xi_{1})_{\widetilde{q}_{1}}; \ldots;  (\xi_{N},\Xi_{N})_{\widetilde{q}_{N}}$.

\textit{Proof:}
From (\ref{pim}) and (\ref{LMGF}) and applying  the binomial expansion \cite[Eq. (1.111)]{grad} in (\ref{fr}), the outage probability  $\widetilde{{ \Pi}}(\rho, N)$ follows  from \cite[Definition A.1]{mathai} after some manipulations.

To better understand the trade-off between $M$ and $N$, a simpler upper bound expression on the outage probability in (\ref{poutd})  is obtained from (\ref{poutL}) and (\ref{fr}) as
\begin{equation}
 \widetilde{{\bar \Pi}}(\rho, N)\!\overset{(a)}{=}\!M\!\int_{0}^{1}(1-x)^{M-1}{\bar \Pi}\left(\frac{\rho}{(d R \sqrt{x})^{-\alpha}}, N\right)dx,
\end{equation}
where $(a)$ follows from letting $x=(r/R)^{2}$  and  applying \cite[Eq. (2.53)]{mathai}, thereby  evolving to (\ref{PLBM}) shown at the top of the next page.
\begin{figure*}
\begin{eqnarray}
\widetilde{ {\Pi}}(\rho, N)\leq\widetilde{{\bar \Pi}}(\rho, N)&=&\tau~\Gamma(M+1) \nonumber\\ && \!\!\!\!\!\!\!\!\!\!\!\!\!\!\!\!\!\!\!\!\!\!\!\!\!\!\!\!\!\!\!\!\!\!\!\!\!\!\!\!\!\!\!\!\!\!\!\!\!\!\!\!\!\!\!\!\!\times {\rm H}_{\sum_{i=1}^{N}{\widehat p}_i+2, \sum_{i=1}^{N}{\widetilde q}_i+2}^{\sum_{i=1}^{N}{\widetilde m}_i,\sum_{i=1}^{N}{\widehat{n}}_i+2}\left[\!\left(\frac{ \sqrt{\rho_t}}{ \left(R d\right)^{-\frac{\alpha}{2}} N}\right)^{N} \prod_{i=1}^{N}{\widetilde{c}}_{i} \! \left|\!\begin{array}{ccc} (0,N \frac{\alpha}{4}),  (1,1),(\delta_{j},\Delta_{j})_{{\widetilde p}_j, j=1:N} \\ (\xi_{j},\Xi{j})_{\widetilde{m}_{j}, j=1:N},(0,1),(\xi_{j},\Xi{j})_{\widetilde{q}_{j}, j=1:N}, (-M ,N \frac{\alpha}{4})\end{array}\right. \right].
 \label{PLBM}
\end{eqnarray}
\hrulefill
\end{figure*}
Similar to (\ref{asymout}), the asymptotic outage probability of BPP distributed RIS-assisted networks  is obtained by evaluating  the residue at the points $\zeta_l = (\zeta_1, \ldots, \zeta_N)$, where $\zeta_l=\underset{j=1,\ldots, \widetilde{m_l}}{\min}\left\{\frac{\xi_{lj}}{\Xi_{lj}}\right\}$. Then, after some manipulations, we  obtain
\begin{eqnarray}
\widetilde{{ \Pi}}(\rho, N)&\approx&  \!\!\!\!\! \frac{\tau \Gamma(M+1) \Gamma\left(1+\frac{\alpha}{4}\sum_{i=1}^{N}\zeta_i\right) }{\Gamma\left(1+\sum_{i=1}^{N}\zeta_i\right)\Gamma\left(1+\frac{\alpha}{4}\sum_{i=1}^{N}\zeta_i+M\right)} \nonumber \\ && \!\!\!\!\!\!\!\!\!\!\!\!\!\!\!\!\!\!\!\!\!\!\!\!\!\!\!\!\!\!\!\!\times\left(\prod_{i=1}^{N} \Theta_i(\zeta_i)\left(\frac{{\widetilde{c}}_{i} }{(d R) ^{\frac{-\alpha}{2}}}\right)^{\zeta_i}\right)\rho_t^{\frac{\sum_{i=1}^{N}\zeta_i}{2}}.
\label{asm1}
\end{eqnarray}
The result in (\ref{asm1})  illustrates that the outage probability is a
monotonically increasing function of the serving distance $d$ and $R$ (Note that $\zeta_i>0$, $i=1,\ldots,N$). Moreover, (\ref{asm1}) shows that increasing $M$ provide sustainable performance gain by considerably reducing the outage probability. 

\textit{Corollary 6  (RIS Deployment Scaling Laws):}
For RIS-assisted communication networks with $M$ spatially distributed RIS, the outage probability has the following scaling law  
\begin{eqnarray}
 {\Pi}(\rho, N)&\underset{M\rightarrow\infty}{\leq}&\tau~ \nonumber\\ && \!\!\!\!\!\!\!\!\!\!\!\!\!\!\!\!\!\!\!\!\!\!\!\!\!\!\!\!\!\!\!\!\!\!\!\!\!\!\!\!\!\!\!\!\!\!\!\!\!\!\!\! \times {\rm H}_{\sum_{i=1}^{N}{\widehat p}_i+2, \sum_{i=1}^{N}{\widetilde q}_i+1}^{\sum_{i=1}^{N}{\widetilde m}_i,\sum_{i=1}^{N}{\widehat{n}}_i+2}\Bigg[\!\left(\frac{ \sqrt{\rho_t}}{ \left(R d\right)^{-\frac{\alpha}{2}} M ^{\frac{\alpha}{4}}N}\right)^{N} \prod_{i=1}^{N}{\widetilde{c}}_{i} \!\nonumber \\ && \!\!\!\!\!\!\!\!\!\! \left|\!\begin{array}{ccc} (0,N \frac{\alpha}{4}),  (1,1),(\delta_{j},\Delta_{j})_{{\widetilde p}_j, j=1:N} \\ (\xi_{j},\Xi{j})_{\widetilde{q}_{j}, j=1:N},(0,1)\end{array}\right. \Bigg].
 \label{PLBMM}
\end{eqnarray}

\textit{Proof:}
The result follows from using the  representation of the involved Fox's H-function in terms of Mellin-Barnes integral in (\ref{PLBM}), then applying  $\underset{M\rightarrow\infty}{\lim }\frac{\Gamma(M+1)}{\Gamma(1+M-N \frac{\alpha}{4})} =M^{-\frac{N\alpha}{4}s}$. Finally, recalling \cite[Eq. (1.2)]{mathai} yields the scaling law of the outage probability as in (\ref{PLBMM}). \\
In the same way as the asymptotic outage  in (\ref{outlb}), we can easily prove that in far-field RIS-assisted communications, the array gain  preserves its quadratic scaling with $N$, while  grows with $M^{\alpha/2}$, where $M$ is the RISs number. This implies that RISs benefit
more from increasing their elements number,  than increasing their density in the network. It is interesting to note that  when $\alpha$ increases, which is the  case of heavily shadowed environments with ($3<\alpha\leq5$) and obstructed in building ($4<\alpha\leq6$) environments, although the RIS network benefits more form increasing $M$,  it does not mean that it will achieve a higher SNR when $M$ is large.

\section{Numerical Results}
In this section, we investigate the performance of the RIS-aided networks and verify our analytical results for the
outage probability and ergodic capacity by Monte-Carlo simulation. The multivariate Fox's H-function in all the previously obtained
expressions is numerically evaluated using an efficient portable implementation code  provided in \cite{Lei},\!\cite{hmulti}.
Unless otherwise stated, the SNR threshold is set to $\rho=0$ dB.

Fig.~1 shows the outage probability vs the average transmit SNR for several values of $N$ under Nakagami-m fading with $m_i^{g}>m^{h}_i$, $i=1,\ldots,N$, with $\underset{i=1,\ldots,N}{\min}\{m^{h}_i\}=0.5$.  Specifically, the proposed outage
exact expression in (\ref{pout}) and its high SNR counterpart in (\ref{ounak}) are
plotted together with the exact Monte-Carlo simulations.  Several observations are gained:
i) Our analytical results in (\ref{pout})  exactly match with
the simulation results, which confirms the accuracy of our
analysis; ii) For different RIS elements count $N$,  we
notice that the outage decreases at a rate of $\rho_L^{-N}$, where $\rho_L$ is the average transmit SNR.  This has been analytically proved in (\ref{ounak}); iii) The figure shows that the total channel gain grows as $N^{2}$ with the RIS, which is consistent with the respective far-field approximations.
\begin{figure}[h]
\centering
\includegraphics[scale=0.35]{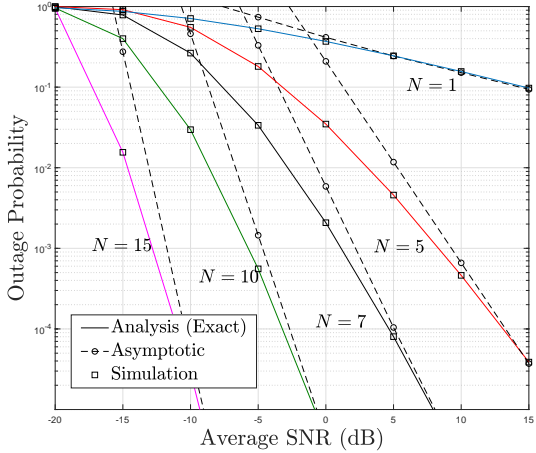}
\caption{ The outage probability vs the average transmit SNR in Nakagami-$m$ fading.}
\end{figure}

Fig.~2 shows the outage probability vs the average SNR over $\alpha$-$\mu$ fading evaluated using  Table I.
 For a given $N$, the outage probability decreases
with $\ln(\rho_L)^{2N-1}\rho_L^{-\frac{N \alpha \mu}{2}}$ in i.i.d. $\alpha$-$\mu$ fading which confirms Corollary 2. However in i.ni.d. $\alpha$-$\mu$ fading the outage probability decreases at a rate of $\rho_L^{-\frac{\underset{i=1,\ldots,N}{\min}{\alpha_i \mu_i}}{2}}$, which corroborates (\ref{asymout}).

\begin{figure}[h]
\centering
\includegraphics[scale=0.35]{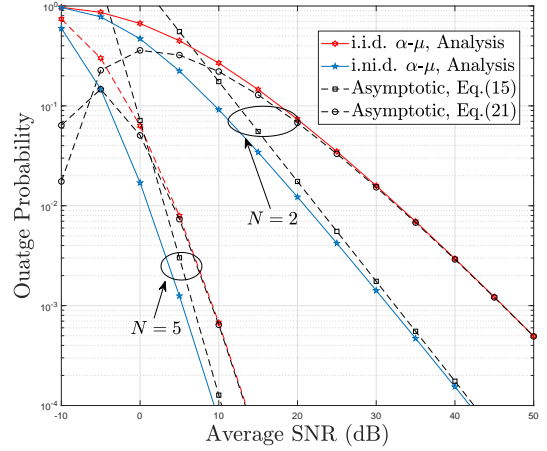}
\caption{ The outage probability vs the average transmit SNR in $\alpha$-$\mu$ fading.}
\end{figure}
Fig~3 illustrates the ergodic capacity of RIS-assisted communications over Nakagami-$m$ fading  for
different reflective RIS elements number $N$. Importantly, our analytical results in Table II exactly match with
the simulation results, which confirms the accuracy of our
analysis. Fig.~3 demonstrates that increasing the RIS elements count definitely benefits the ergodic capacity, while the capacity improvement  diminishes as the number of elements grows large.

\begin{figure}[h]
\centering
\includegraphics[scale=0.35]{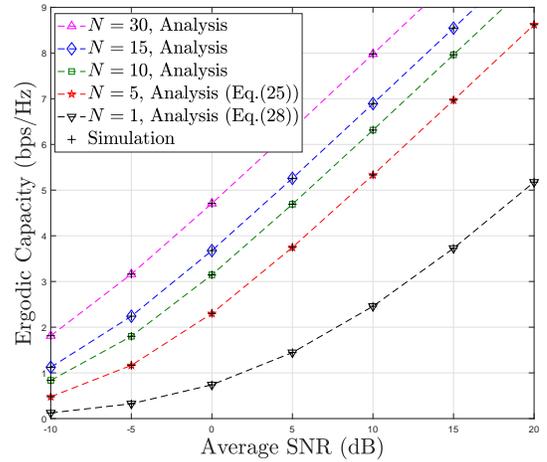}
\caption{The average capacity versus the average transmit SNR in Nakagami-$m$ fading.}
\end{figure}

Fig.~4 depicts the ergodic capacity of a singe-element RIS-assisted communication using (\ref{cap1}).   In the legend, we have identified some particular fading distribution cases that simply stem from the general Fox's H function fading model.   The latter, includes as special cases generalized-$\cal K$ with heavy ($\kappa=0.5$) and moderate ($\kappa=1.5$) shadowing,  Nakagami-$m$, Rayleigh ($m=1$), and $\alpha$-$\mu$ fading models, to name a few (see Table II, \cite{FoxH}). It is worthy to  note that such
result is totally new  and generalises and unifies all previous results pertaining to single-element RIS performance (\!\cite{ray} and references therein).

\begin{figure}[h]
\centering
\includegraphics[scale=0.35]{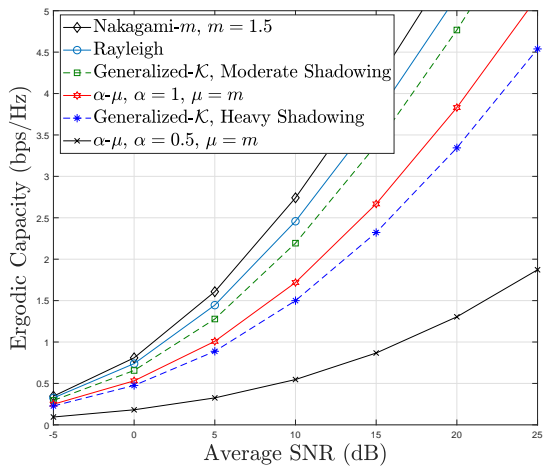}
\caption{The average capacity versus the average transmit SNR when $N=1$.}
\end{figure}

In fig.~5,  we investigate the performance of  RIS deployment using (\ref{poutd})  under different number of RISs $M$
and pathloss exponent $\alpha$.  It is observed that both $N$ and $M$ improve the system performance. Yet, interestingly,
it is more beneficial to assemble more elements into fewer RISs, in
order to maximise the RIS-enabled network outage performance. The performance gap reduces
asymptotically but will not vanish, which is consistent with Corollary 6.
This is mainly due to the prominent SNR quadratic scaling with $N$, thus resulting
in a significant throughput  improvement for nearby users, whose performance
gains compensates the generally increased outage
areas with fewer RISs.

\begin{figure}[h]
\centering
\includegraphics[scale=0.35]{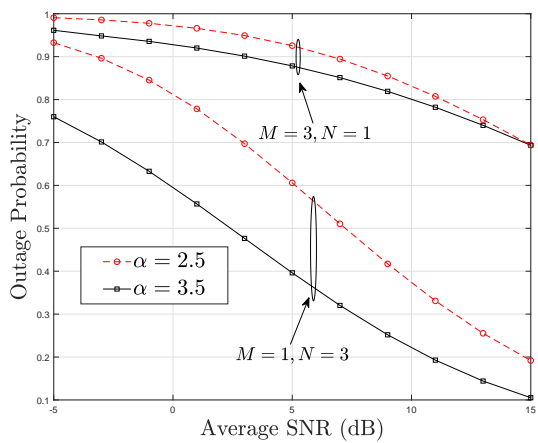}
\caption{Spatial outage  in Nakagami-$m$ fading with $m=0.5$ and under different
path-loss exponent $\alpha$ and number of RIS $M$ with  $M N = 3$
and $R = 50$m.}
\end{figure}

\section{Conclusion}
It has been so far  widely admitted that the exact performance of RIS-based communications for arbitrary number of elements $N$ and general fading distribution is rather untractable due to the inherent intricacy of the subject treatment \cite{basar1}-\!\!\cite{ray}. In this paper, we have successfully tackled the problem  by  providing the exact characterisation of outage and ergodic capacity performance of RIS-based communication while remarkably incorporating prominent and generalized fading distributions. Moreover, asymptotic analysis has been conducted for high
SNR regime. Our analysis unveils two scaling rates for outage probability both governed by the worst fading multiplied by the RIS elements count $N$.     Capitalizing on the developed statistical framework, we have also characterized the spatial
performance of randomly deployed  multiple
RISs-based communications. We show that there is a great  potential to
improve  the outage performance and thereby the capacity  when fewer RISs are deployed each
with more reflecting elements. The obtained results could provide helpful guidance for the
network deployment and application of RIS technology in future wireless networks.

\end{document}